\documentclass[prl,reprint,superscriptaddress]{revtex4-1}
\usepackage{graphicx}
\usepackage{amsfonts}
\usepackage{amssymb,amsmath}
\setcitestyle{super}

\begin{document}
\title{Induced transparency by coupling of Tamm and defect states in tunable terahertz plasmonic crystals}
\author{Gregory C. Dyer}
\email{gcdyer@sandia.gov}
\affiliation{Sandia National Laboratories, P.O. Box 5800, Albuquerque, NM 87185 USA}
\author{Gregory R. Aizin}
\affiliation{Kingsborough College, The City University of New York, Brooklyn, New York 11235 USA}
\author{S. James Allen}
\affiliation{Institute for Terahertz Science and Technology, UC Santa Barbara, Santa Barbara, California 93106 USA}
\author{Albert D. Grine}
\author{Don Bethke}
\author{John L. Reno}
\author{Eric A. Shaner}
\thanks{Notice: This manuscript has been authored by Sandia Corporation under Contract No. DE-AC04-94AL85000 with the U.S. Department of Energy. The United States Government retains and the publisher, by accepting the article for publication, acknowledges that the United States Government retains a non-exclusive, paid-up, irrevocable, world-wide license to publish or reproduce the published form of this manuscript, or allow others to do so, for United States Government purposes.}
\affiliation{Sandia National Laboratories, P.O. Box 5800, Albuquerque, NM 87185 USA}

\begin{abstract}
Photonic crystals and metamaterials have emerged as two classes of tailorable materials that enable precise control of light. Plasmonic crystals, which can be thought of as photonic crystals fabricated from plasmonic materials, Bragg scatter incident electromagnetic waves from a repeated unit cell. However, plasmonic crystals, like metamaterials, are composed of subwavelength unit cells. Here, we study terahertz plasmonic crystals of several periods in a two dimensional electron gas. This plasmonic medium is both extremely subwavelength ($\approx \lambda/100$) and reconfigurable through the application of voltages to metal electrodes. Weakly localized crystal surface states known as Tamm states are observed. By introducing an independently controlled plasmonic defect that interacts with the Tamm states, we demonstrate a frequency agile electromagnetically induced transparency phenomenon. The observed 50\% {\it in-situ} tuning of the plasmonic crystal band edges should be realizable in materials such as graphene to actively control the plasmonic crystal dispersion in the infrared.
\end{abstract}
\maketitle

\begin{figure*}
\includegraphics[scale=1]{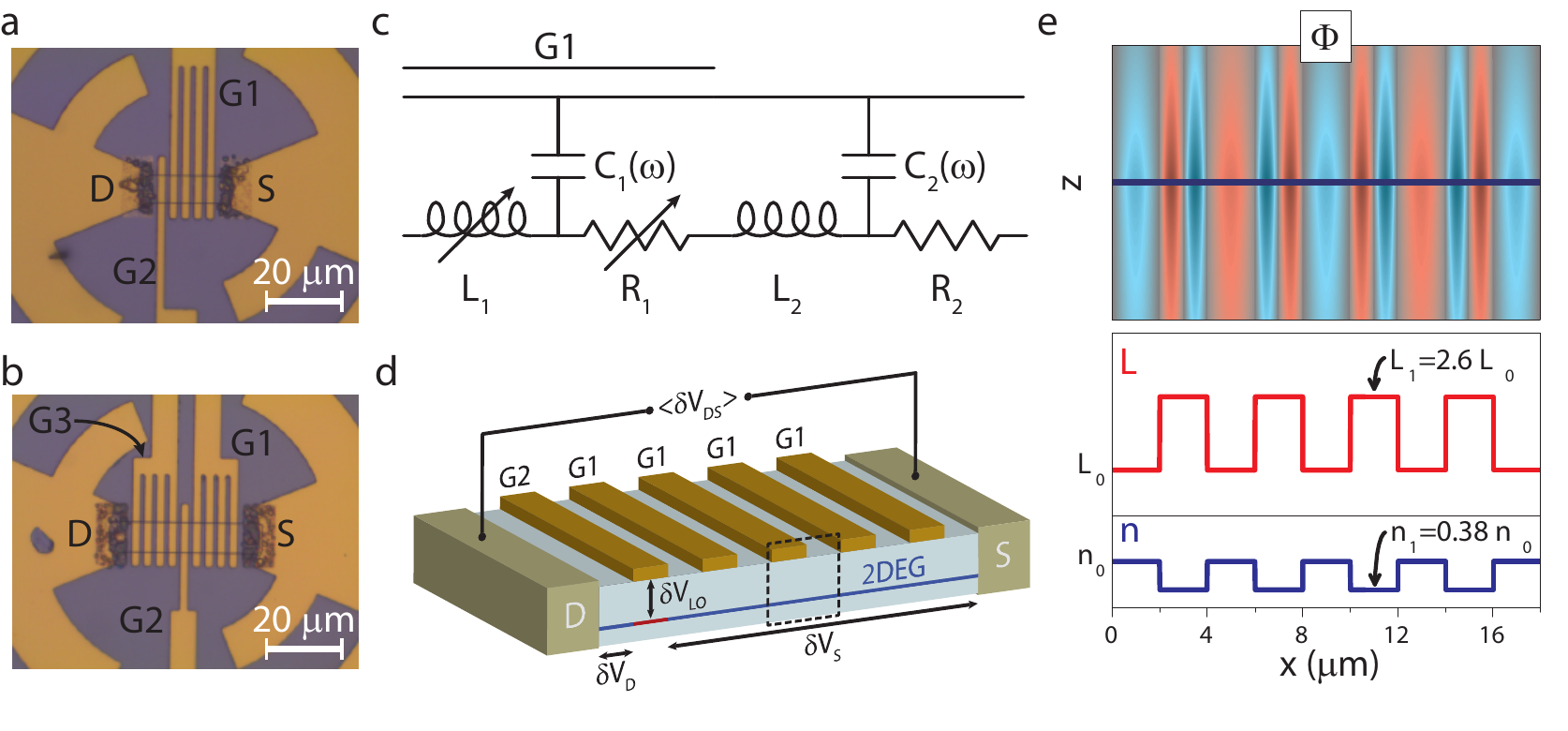}
\caption{{\bf Integrated Plasmonic Crystal Structures.} Images at the broadband antenna vertex of (a) Sample A and (b) Sample B are shown with gates (G1 and G2 for both samples, G3 for Sample B only), source (S) and drain (D) terminals labeled. (c) Equivalent distributed circuit for the PC unit cell with gated and ungated region elements below G1. (d) Diagram of Sample A illustrating the coupling of fields to the plasmonic mixer induced below G2 with the depleted region shown in red and the 2DEG indicated by the blue dashed line. The photovoltage is measured between D and S terminals. (e) The distribution in the x-z plane of the potential $\Phi$ near the plane of the 2DEG is plotted for a resonance ($L_1=2.6 L_0$, $\nu=456$ GHz) of the PC formed between G2 and S. The plane of the 2DEG is indicated by a dark blue line. Shown below are the equilibrium 2DEG inductance $L$ (red) and density $n$ (blue) of the PC.}
\end{figure*}

Photonic band gaps\cite{Yablonovitch1991a}, strong light-matter interaction\cite{Yoshie2004}, slow light\cite{Baba2008}, and negative refractive index\cite{Berrier2004} arise in photonic crystal\cite{Yablonovitch1987,John1987} structures due to Bragg scattering of electromagnetic waves from a repeated unit cell.  However, the electromagnetic properties of photonic crystals engineered from bulk semiconductors, metals, and dielectrics generally are weakly tunable, if at all. Material systems such two dimensional electron gases (2DEGs) embedded in semiconductors\cite{Allen1977,Dyakonov1996} and graphene\cite{Ju2011,Yan2012,Grigorenko2012} offer a substantially more flexible electromagnetic medium. These plasmonic materials can both be lithographically patterned and electronically tuned, giving rise to a variety of subwavelength plasmonic devices that may be broadly controlled via an applied DC electric field. When a periodic structure is engineered from these systems, plasmonic band structure can be realized\cite{Mackens1984,Muravev2008,DyerPRL2012,Andress2012}.  The 2DEG and graphene thus provide a platform for the exploration of widely tunable plasmonic band gap structures.

Subwavelength plasmonic media that utilize a 2DEG formed at a GaAs/AlGaAs interface are the central focus of this article. Similar to the $\omega-q$ plasmon dispersion in graphene, the 2DEG plasmon dispersion depends explicitly upon both the plasmon wavevector and the AC conductivity of the medium. An effective methodology to describe plasma excitations in a 2DEG is that of an `$LC$' plasmonic resonator\cite{Burke2000}. Here $L$ is the field effect tunable kinetic inductance of the 2DEG. The 2DEG capacitance can be introduced as 
\begin{equation}
C=2\epsilon_{eff} q
\end{equation}
where $\epsilon_{eff}$ is the effective permittivity of the embedded 2DEG and $q$ is the plasmon wavevector\cite{DyerPRL2012,Rana2008,Staffaroni2012,Aizin2013}. In high mobility 2DEG materials at microwave and THz frequencies, underdamped `$LC$' plasma resonances are supported, allowing for propagation lengths on the order of tens of micrometers or plasmon wavelengths.

The introduction of spatial periodicity to a 2DEG produces a plasmonic crystal (PC) where the 2DEG is a coherent plasmonic medium. Though a PC is physically more similar to a photonic crystal than a metamaterial, the unit cell of a 2DEG PC is deeply subwavelength. These tunable plasmonic materials are conceptually related to certain periodic plasmonic nanostructures\cite{Shvets2004}, resonant microwave metalenses\cite{Lemoult2010}, and acoustic metamaterials\cite{Lemoult2011}. In such systems, band structure results from incident waves interacting with spatially periodic subwavelength resonant scatterers\cite{Liu2000,Davanco2007}.

\section{2DEG plasmonic crystals}

In this article, we examine the complex interplay between surface states known as Tamm states\cite{Tamm1932}, plasmonic defect (PD) modes, and the PC band structure in tunable THz plasmonic band gap devices engineered from a GaAs/AlGaAs 2DEG. The studied PC devices, Samples A and B, are pictured in Fig. 1a,b. Both are integrated at the vertex of an antenna having bandwidth from 100 GHz to 1 THz. These plasmonic structures are based upon a four period PC formed below gate G1 with an adjacent independently controlled plasmonic defect (PD) \cite{Shaner2006,Davoyan2012} controlled by gate G2. A PD is induced when G2 is tuned to a different voltage than G1. While these plasmonic band gap structures are nearly 100 times smaller than the free space wavelength of THz radiation, they cannot be considered an effective medium because the wavelengths of the THz plasmons tightly confined to the 2DEG are comparable to the size of the PC unit cell\cite{Smith2005,Simovski2007}.

The adjacent regions of 2DEG in this system can be represented as sequential distributed plasmonic `$RLC$' transmission line elements, where $C$ is defined in Eq. 1. It is convenient to treat the PC unit cell as illustrated in Fig. 1c using this equivalent circuit approach. In Fig. 1d the PC unit cell is indicated by a dashed box in the cross-sectional illustration of Sample A. The kinetic inductance $L$ and resistance $R$ of the equivalent distributed circuit are explicitly defined through the Drude conductivity of the 2DEG, $\sigma (\omega)^{-1}=R+i \omega L$\cite{Burke2000,YoonHam2012}  ({\it Supplemental Materials, Section I}).

\begin{figure*}
\includegraphics[scale=1]{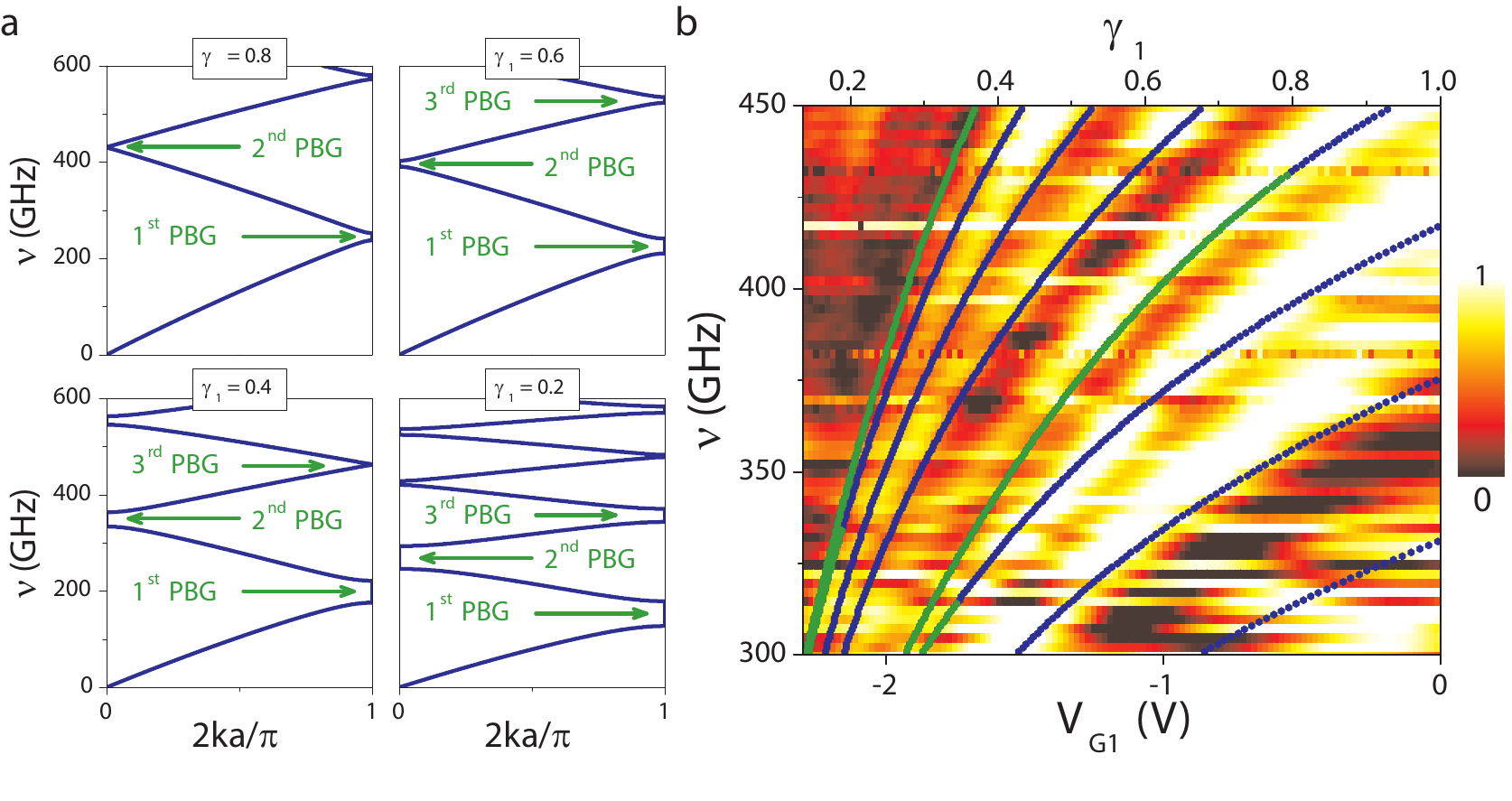}
\caption{{\bf Tunable Plasmonic Crystal Spectrum.} (a) The frequency-wavevector dispersion of the plasmonic crystal based on the unit cell in Fig. 1c is plotted for $\gamma _1$=0.8, 0.6, 0.4 and 0.2. Only positive wavevectors are shown. (b) The self-normalized PC photovoltage spectrum of Sample A as a function of $V_{G1}$ and frequency is plotted.  Eight calculated PC modes are shown in blue, with sections of the modes that are found in the infinite crystal band gap indicated in green.}
\end{figure*}

\begin{figure}
\includegraphics[scale=1]{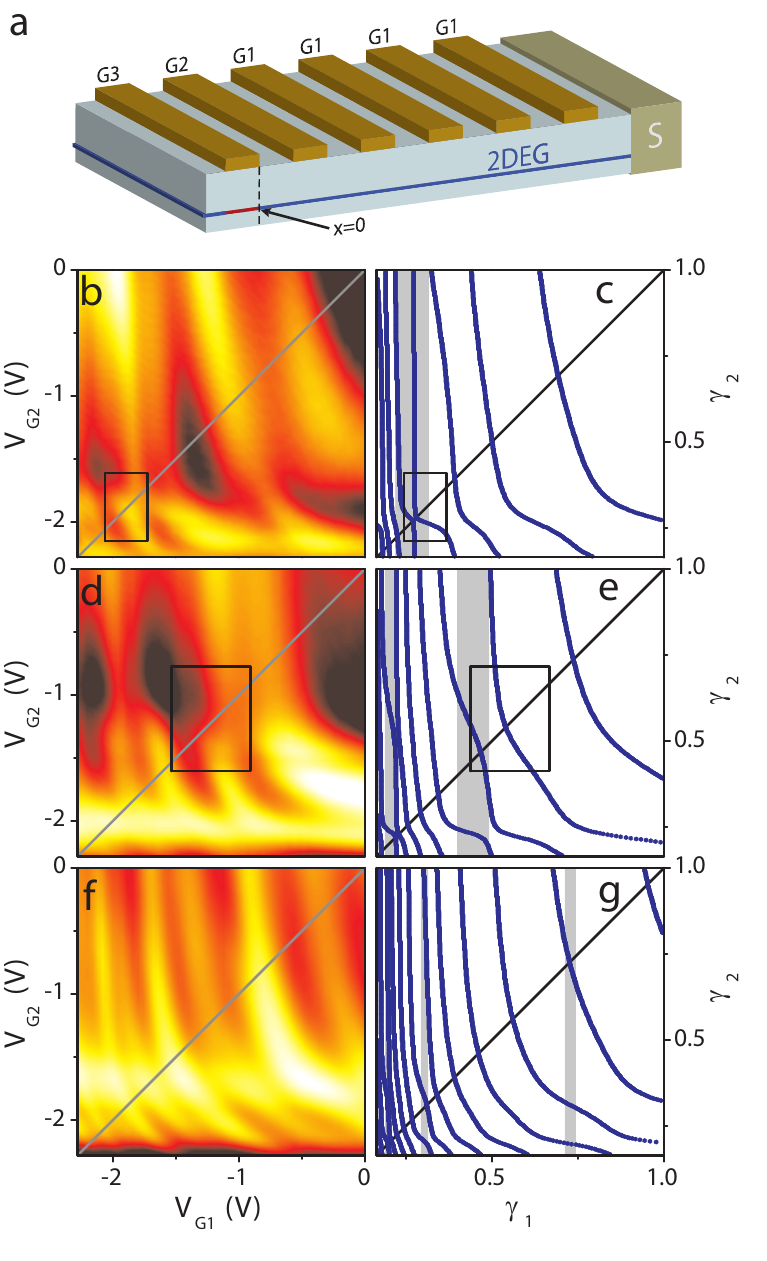}
\caption{{\bf Tamm States in Plasmonic Crystal-Defect Structures.} (a) A schematic of Sample B configured such that a four period PC, tuned by G1, with adjacent PD, tuned by G2, is formed between S and G3. The plasmonic detection region is indicated by red dashes. Plots of the plasmonic photovoltage spectra of Sample B as a function of $V_{G1}$ and $V_{G2}$ are shown for (b) 302.5, (d) 363.0 and (f) 420.5 GHz excitation frequencies.  The calculated band gaps (grey) of the infinite PC and the PC-PD system modes (blue) are plotted for (c) 302.5, (e) 363.0 and (g) 420.5 GHz. }
\end{figure}

\begin{figure}
\includegraphics[scale=1]{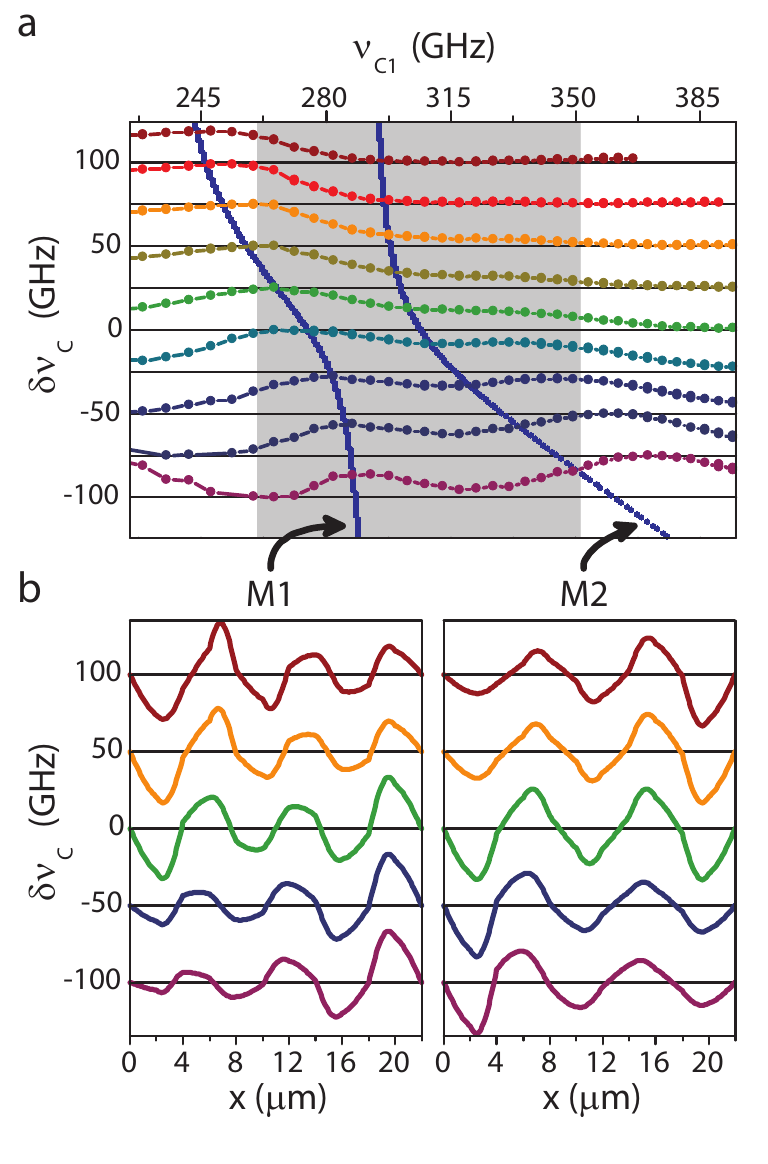}
\caption{{\bf Induced Transparency in the First Plasmonic Band Gap.} (a) The normalized photovoltage as a function of the PC characteristic frequency is shown for a 210.0 GHz excitation. Each curve corresponds to a different detuning of the PD characteristic frequency relative to that of the PC unit cell from -100 to +100 GHz in 25 GHz steps.  The calculated modes are plotted in blue with band gaps shaded in grey. The arrows highlight modes denoted M1 and M2. (b) The spatial voltage distributions in the plane of the 2DEG for M1 and M2 with several detunings are plotted and highlight the interaction of PD and Tamm states under 210.0 GHz excitation.}
\end{figure}

Tuning of the gate voltages $G1$ and $G2$ controls the 2DEG inductance and resistance, $L_j,R_j \propto 1/ \gamma_j$. Here $\gamma_j$ defines the normalized 2DEG density in terms of the threshold voltage $V_{th}$ (where $n_{2D}\rightarrow 0$) and the applied gate voltages $G_j$ such that $\gamma _j\equiv (V_{th}-V_{Gj})/V_{th}$. In terms of the PC unit cell diagrammed in Fig. 1c, the applied gate voltage $V_{G1}$ controls the inductance $L_1$ and resistance $R_1$. The ungated region of 2DEG has constant distributed inductance $L_0$ and resistance $R_0$.

In Fig. 1e, the electrostatic potential $\Phi$ of a plasma wave with a frequency of $\nu=456$ GHz is illustrated in a four-period PC. The potential $\Phi$ is shown around the plane of the 2DEG. Also indicated are the spatially periodic equilibrium 2DEG kinetic inductance $L$ and density $n$ of the four-period PC. Here a 2 $\mu$m plasmon wavelength below G1 is evident for the 456 GHz excitation with free space wavelength of 658 $\mu$m, or about $\lambda/300$.

\section{Active control of plasmonic band structure}

To develop intuition concerning the tunability of the PC, we first consider the bulk plasmonic band structure of a PC having an infinite number of periods. The crystal dispersion of the infinite PC, plotted in Fig. 2a for several values of $\gamma _1$, was calculated from the 1D Kronig-Penney model\cite{Aizin2013} for the unit cell shown in Fig. 1c,
\begin{multline}
cos(2k_B a)=cos(q_1a) cos(q_0a) \\
-\frac{1}{2} \Biglb( \frac{Z_1}{Z_0}+\frac{Z_0}{Z_1} \Bigrb) sin(q_1a)sin(q_0a).
\end{multline}
Here $k_B$ is the Bloch wavevector, $q_j$ and $Z_j$ are the plasmon wavevector and characteristic impedance, respectively, of the two elements forming the unit cell, and $a$ is the length of the $j^{th}$ section of 2DEG ({\it Supplemental Materials, Section II}). With $\gamma _1$=0.8, minigaps begin to emerge in the 2DEG plasmon dispersion relation that is folded into the first Brillouin zone. As $\gamma _1$ further decreases, full plasmonic bandgaps are evident and the entire band structure shifts downward in frequency.

A four-period PC between S and G2 can be induced in Sample A, shown in Fig. 1a, by biasing the defect gate G2 past its threshold voltage as illustrated in Fig. 1d. The PC formed below G1 delivers a plasmonic signal to a rectifying detection element under G2, generating a photovoltage $\langle \delta V_{DS} \rangle$ as explained in the Methods Summary. By tuning G1 and frequency as plotted in Fig. 2b, the measured photoresponse maps the plasmonic spectrum of Sample A. The bright regions correspond to the resonant excitation of plasma modes in the system ({\it Supplemental Materials, Section II}). For comparison, resonant modes calculated for the four-period PC using a transfer matrix formalism\cite{DyerPRL2012,Aizin2013} ({\it Supplemental Materials, Section I}) are shown by solid lines. These eight resonances are associated with the second and third allowed bands, and the five lowest order of these resonances are resolved experimentally. Above 350 GHz and $V_{G1}<-2.00 \: V$, discrete modes cannot be resolved due to the relatively shorter plasmon coherence length in this regime ({\it Supplemental Materials, Section III}).

Each unit cell of the four-period structure would have an identical resonant frequency in isolation. However, the coherence of the plasma wave across the PC lifts this four-fold degeneracy, resulting in the formation of four state bands. Alternately, this can be considered a Fano-type system where interference between a continuum, the allowed bands of the PC, and the discrete modes of the cavity formed between S and G2 produces groups of four asymmetric resonances\cite{Davanco2007}. These bands of states are demonstrated in Fig. 2a to tune a minimum of 50\% in frequency from 300 GHz to 450 GHz, though in principle the shift of the band gap can be well in excess of 100\%.

The first three experimentally observed resonances in Fig. 2b moving from the lower right-hand corner (300 GHz and $V_{G1}=0.00 \: V$) towards the upper left-hand corner are positioned in an allowed infinite PC band. These band states are highlighted in blue. However, the fourth mode associated with this band is largely found in the second band gap of the infinite PC, where it is highlighted in green. All regions of modes highlighted in green are in the predicted band gaps corresponding to those labeled in Fig. 2a, including a pair near $V_{G1}=-1.80 \: V$ and 300 GHz. These modes could potentially represent plasmonic Tamm states\cite{Tamm1932,Aizin2013}. Tamm states, weakly localized crystal surface states with a complex rather than purely real Bloch wavevector ($Im[k_B] \neq 0$), are generally found in a band gap and have been experimentally demonstrated previously only in a few electronic\cite{Ohno1990}, photonic\cite{Goto2008}, and hybrid optical-plasmonic\cite{Sasin2008} systems. In contrast to surface states resulting from dislocations and impurities, Tamm states form at the ideal termination of a lattice and are most easily isolated in tailorable structures like semiconductor superlattices and photonic crystals. Though the results shown in Fig. 2 are indicative of Tamm state formation, additional direct measurements are needed to justify this hypothesis.

\section{Evidence of plasmonic Tamm states}

To search for Tamm states in this system, Sample B, pictured in Fig. 1b, was studied with gate G3 biased beyond its threshold voltage. Here the last stripe of depleted 2DEG below G3 operates as a rectifying detector as illustrated in Fig. 3a ({\it Supplemental Materials, Section II}). This configuration of Sample B provides a means to study the strong coupling between the four-period PC below G1 and a the PD\cite{Shaner2006,Davoyan2012,Muravev2012} defined under G2. The photoresponse to excitation frequencies of 302.5, 363.0 and 420.5 GHz as G1 and G2 are independently tuned is shown in Fig. 3b,d,f, respectively. As in Fig. 2a, the bright regions represent plasma resonances. In Fig. 3c,e,g the calculated PC-PD structure plasmon modes, blue lines, are plotted for these same frequencies as a function of normalized 2DEG densities $\gamma _j$ corresponding to the experimentally applied gate voltages. The band gaps of the infinite PC where the Bloch wavevector of experimentally observed resonances has a non-zero imaginary component are indicated in grey.

An understanding of the relationship between the PC and PD states emerges when considering Fig. 3b-g. The four period PC modes are tuned by G1 and appear as vertical features when the PD is not resonating. PD resonances are modes controlled by G2 and appear as bright horizontal lines where sequential anticrossings occur. When $V_{G1}=V_{G2}$, the PD may be viewed as a fifth identical unit cell that is appended to the four-period PC under G1. This condition is satisfied along the diagonal lines in Fig. 3b-g where a fifth discrete mode is added to each band by a higher order mode moving into a lower energy band after traversing the infinite PC band gap.

The black boxes in Fig. 3b-e highlight repelled crossings between two states in or near the second band gap along the diagonal line $V_{G1}=V_{G2}$. A repelled crossing is found along this diagonal if and only if a PD and a Tamm state of the same order are mutually coupled. Because an isolated PD state is localized, it must enter into a band gap before joining a five-state band as an orthogonal PC mode. Only localized states are found in the infinite PC band gap. A Tamm state bound near the Ohmic contact is the only viable candidate to cross with a PD mode along the line $V_{G1}=V_{G2}$.

There are several additional distinctions between crossings of the PD modes with purely real-valued Bloch wavevector PC states and with complex Bloch wavevector Tamm states. Modes in the allowed bands are completely delocalized and therefore are widely separated as they come into resonance with a PD mode. In Fig. 3f,g under 420.5 GHz excitation where the band gaps largely vanish, none of the PC states becomes localized and for any fixed choice of $V_{G1}$ ($\gamma _1$) the resonances are widely spaced in $V_{G2}$ ($\gamma _2$). In contrast, due to the localization of Tamm states, both their coupling with the PD and the size of the PD-Tamm splitting are comparatively smaller. This leads to the the close approach of the modes in the black boxed regions of Fig. 3b-e.

\section{Plasmon-induced transparency through coupled localized resonances}

As the energy of a PD mode approaches that of the Tamm state by tuning $V_{G2} \rightarrow V_{G1}$, an analogy may be drawn to electromagnetically induced transparency (EIT)\cite{Zhang2008,LiuNM2009,Tassin2009} provided several conditions are satisfied. The description of EIT in classical systems is that of two coupled oscillators, one of which is `bright' and and the other `dark' with respect to incident radiation. In this picture, the PD and Tamm states then must be coherently coupled and resonate at the same frequency, conditions that are satisfied empirically in this plasmonic system. More importantly, there must be an asymmetry in both the external coupling and the damping rates of the coupled plasmonic oscillators. The `dark' resonance must be weakly externally coupled and have a higher quality factor than the `bright' resonance. In high quality 2DEG plasmonic systems, weak coupling to an external excitation implies a higher `Q' resonance because radiative damping is the dominant dissipation mechanism of the plasmon ({\it Supplemental Materials, Section III}).

This plasmonic EIT-like effect is studied through analysis of the PC-PD modes in and near the first band gap. While it is common to sweep frequency in order to map the interaction of strongly coupled resonators\cite{Luk2010}, our approach of tuning resonators {\it in-situ} with fixed excitation frequency yields the same information. In Fig. 4a the photoresponse of Sample B with an excitation frequency of 210.0 GHz is shown. Here the characteristic frequency $\nu _{C1}$ of the PC unit cell is swept for different detunings $\delta \nu _C$ of the PD relative to PC unit cell. The characteristic plasma frequencies
\begin{equation}
\nu _{Cj}=\beta_C/2 \pi \sqrt{L_{j} C_{j}}
\end{equation}
are defined for the PC unit cell ($j=1$) and PD ($j=2$) using the fundamental wavevector $\beta _C \equiv \pi / a$ of the plasmon mode confined below an $a=2 \: \mu m$ wide gate finger.  In fact, the characteristic frequencies $\nu _{Cj}$ are the fundamental plasma frequencies of the isolated PC unit cell and PD. The detuning of the PD characteristic frequency from that of the PC unit cell is given by $\delta \nu_C=\nu _{C2}-\nu_{C1}$. These definitions of $\nu _{Cj}$ parameterize the strong coupling behavior of the PC-PD system. 

For positive detunings $\delta \nu_C > 50$ GHz, an asymmetric resonance associated with the PD is observed at the edge of the band gap (shaded grey) in Fig. 4a, while the Tamm state appears inert due to its poor coupling to both the incident THz field and the integrated detector.  Here we interpret the PD as a `bright' resonator that under appropriate conditions can couple to and drive the `dark' Tamm state. The PD and Tamm states become strongly interacting when the detuning is reduced, $-50$ GHz $\leq \delta \nu_C \leq +50$ GHz. As the energetic difference between these states decreases, the `bright' PD excites the formerly `dark' Tamm state. A characteristic signature of EIT, a symmetric double peak with a dip at its center, is evident with the detuning set to $\delta \nu_C = -50$ GHz. For negative detunings $\delta \nu_C < -50$ GHz, the PD state shifts to larger characteristic frequencies $\nu _{C1}$ and the Tamm state's amplitude decreases. The calculated modes, shown in blue in Fig. 4a, agree well with the measurements.

The plasmonic spatial distributions in Fig. 4b highlight the evolution of the PD and Tamm states. The distributions of modes M1 and M2, indicated by the arrows linking to Fig. 4a, are plotted for several detunings $\delta \nu_C$. The position corresponds directly to Fig. 3a; the mixer edge is at $x=0$, the PD is located between $2<x<4$, and the PC is found between $6<x<22$. Both the `bright' PD state, seen distinctly with $\delta \nu_C>0$ for M1 and $\delta \nu_C<0$ for M2, and the `dark' Tamm state, observed with $\delta \nu_C<0$ for M1 and $\delta \nu_C>0$ for M2, are weakly localized. For $\delta \nu_C=0$, the crystal is sufficiently short that the spatial overlap of the two weakly localized states lifts their degeneracy. PC modes develop around zero detuning that are analogous to bonding and anti-bonding states. These may also be interpreted as coupled Tamm states formed at opposing edges of the PC\cite{Fowler1933,Shockley1939}.

\section{Conclusions and Outlook}

In this article, we have studied the band structure as well as coupled surface and defect states in tunable PC structures. The engineering of plasmonic resonators {\it in-situ} rather than solely through lithographic tuning of physical geometry opens a previously unexplored avenue for the study of strongly coupled electromagnetic systems\cite{Luk2010}. Both photonic crystals\cite{Baba2008} and coupled resonators\cite{Totsuka2007,Zhang2008} can be harnessed for slow light applications, and tunable plasmonic systems provide a useful degree of freedom for possible slow light devices. While the presented plasmonic devices based upon GaAs/AlGaAs heterostructures are likely limited in both operating temperature ($<$ 77 K) and operating frequency ($<$ 1 THz), both GaN-based 2DEGs and graphene\cite{Engheta2011} hold promise for extending the viable range of the PC-based structures ({\it Supplemental Materials, Section IV}). Recent studies have shown tunable THz plasmons in GaN 2DEGs at 170 K\cite{Muravjov2010} and at room temperature in graphene\cite{Ju2011,Yan2012}, as well as mid-infrared plasmons in graphene nanostructures\cite{Koppen2012,Basov2012}. Finally, we note that manipulation of the localized PC defect and Tamm state field distributions to produce strong field enhancements\cite{Davoyan2012} in 2DEG structures could give rise to a new generation of ultra-sensitive direct and heterodyne THz detectors as well as THz oscillators\cite{SydorukOpEx2012}.

%\bibliography{bib_v8p0}

{\bf Methods Summary} The devices were fabricated using standard contact lithography, metalization deposition, and lift-off techniques from a GaAs/AlGaAs double quantum well heterostructure (Sandia wafer EA1149) with total 2DEG density 4.02 x 10$^{11}$ cm$^{-2}$ at 12 K. The 2DEG is embedded a distance $d$ = 386 nm below the surface of the MBE-grown heterostructure. The periodic gates were designed to have a 4.0 $\mu$m period with 50\% metalization duty cycle, while fabricated dimension differed slightly yielding a 3.8 $\mu$m period with approximately 60\% metalization duty cycle of the 10 $\mu$m wide mesa.  The THz radiation was generated with a Virginia Diodes, Inc. (VDI) microwave frequency multiplication chain and optically coupled through z-cut quartz windows in a closed cycle cryogenic system.  Photoresponse and transport measurements were performed at 75 Hz modulation rate using a Stanford Research 830 lock-in amplifier to measure the voltage between source and drain terminals.

A purely electronic approach was employed to probe the plasmonic band gap structures. Depletion of the 2DEG ($n_{2D}\rightarrow 0$) below a gate allows for a conversion of plasma waves into a measurable photoresponse\cite{Shaner2006,Dyakonov1996} that can have both photoconductive\cite{DyerAPL2012} and photovoltaic\cite{Muravev2012} contributions. In this work, the distributed THz excitation of the device terminals produces a plasmonic homodyne mixing response\cite{Lisauskas2009,Preu2012IEEE,Klimenko2012} measured between the drain (D) and source (S) contacts as illustrated for Sample A in Fig. 1d.  The local oscillator voltage $\delta V_{LO}$ is coupled from G2 to the region of depleted 2DEG below, while the signal coupled to the mixer is the difference between the plasmonic voltages $\delta V_D-\delta V_S$ generated on either side of the depletion region. A non-linear plasmonic mixing mechanism in the region of depleted 2DEG down converts the component of the differential THz signal $\delta V_D-\delta V_S$ in-phase with $\delta V_{LO}$ to a photovoltage $\langle \delta V_{DS} \rangle$ ({\it Supplemental Materials, Section II}).

The threshold voltage $V_{th}$ of both devices was found using standard lock-in techniques to measure transistor channel conductivity. A fit of the conductivity near threshold was extrapolated to determine $V_{th}=-2.67 \: V$ for Sample A and $V_{th}=-2.73 \: V$ for Sample B. Photoresponse measurements presented in the article were  performed with $V_{G2}=-2.80 \: V$ for Sample A and $V_{G3}=-2.80 \: V$ for Sample B. Samples A and B were characterized independently and aligned to the THz source using a raster scan to locate the maximum photovoltage signal at a given excitation frequency. All measurements were performed at T = 8 K.

{\bf Acknowledgments} The work at Sandia National Laboratories was supported by the DOE Office of Basic Energy Sciences. This work was performed, in part, at the Center for Integrated Nanotechnologies, a U.S. Department of Energy, Office of Basic Energy Sciences user facility. Sandia National Laboratories is a multi-program laboratory managed and operated by Sandia Corporation, a wholly owned subsidiary of Lockheed Martin Corporation, for the U.S. Department of Energy’s National Nuclear Security Administration under contract DE-AC04-94AL85000.

{\bf Author Contributions} G.C.D., S.J.A. and E.A.S. conceived of and designed the devices. J.L.R. grew the 2DEG material. E.A.S. and D.B. fabricated and imaged the devices. A.D.G. and G.C.D. assembled the experiment. G.C.D. measured and analyzed the data. G.R.A. and G.C.D. developed the theory and performed the model computations. G.C.D. wrote the manuscript with editorial input from G.R.A. and E.A.S. All authors discussed the results and commented on the paper.

{\bf Additional Information} Supplementary information is is available in the online version of the paper. Reprints and permissions information is available at www.nature.com/reprints. The authors declare no competing financial interests. Correspondence and requests for materials should be addressed to G.C.D.
\end{document}